\documentclass[12pt]{article}
\usepackage{amsmath} 
\usepackage{mathptmx}
\usepackage{iopconf}
\usepackage{graphicx}
\usepackage{bm}
\usepackage{amsthm}

\begin{document}
\title{On the connection between the Schr\"odinger and the Heisenberg pictures
for unbounded operators}

\author{Rafael de la Madrid}

\affil{Departamento de F\'\i sica Te\'orica, Facultad de Ciencias,
Universidad del Pa\'\i s Vasco, E-48080 Bilbao, Spain \\
E-mail: {\texttt{wtbdemor@lg.ehu.es}} }
%URL: {\texttt{http://www.ehu.es/$\sim$wtbdemor}}}

\beginabstract
It is well known that the unboundedness of operators in Hilbert space 
entails domain troubles. It is also well known that most domain troubles can 
be surmounted by extending the Hilbert space to a rigged Hilbert space. In 
this note, we point out another of such troubles, namely the correspondence 
between the Schr\"odinger and the Heisenberg pictures for unbounded operators, 
and sketch the solution of this problem within the rigged Hilbert space.
\endabstract

\section{Introduction}

Quantum Mechanics textbooks show that the Schr\"odinger and the Heisenberg 
pictures are physically equivalent, because they yield the same probability
amplitudes for measuring an observable $A$ in a state $\varphi$. Textbooks,
however, usually omit the fact that for unbounded operators, the
manipulations that lead from the Schr\"odinger to the Heisenberg picture
must be taken with care, due to domain problems. The purpose of this note
is to point out those domain problems and to sketch their solution by
extending the Hilbert space to the rigged Hilbert space. For a class of
potentials, we use a theorem by Hunziker to solve such problems explicitly.

\section{The problem}

Suppose that the algebra ${\cal A}$ of observables of a system consists of
the following operators:
\begin{equation}
     {\cal A}= \{ H, A_1, A_2, \ldots , A_N \}  \, , 
\end{equation}
where $H$ is the Hamiltonian and $A_1$, $A_2$, $\ldots$, $A_N$ 
are the other relevant observables of the system (e.g., position and 
momentum). Those operators are assumed to be self-adjoint on a Hilbert space
${\cal H}$. For simplicity, we restrict ourselves to pure states and to 
observables that do not depend explicitly on time. Then, in the Schr\"odinger
picture, which shall be denoted by the subscript ${\rm S}$, the
observables are kept fixed in time,
\begin{equation}
      A_{\rm S}(t)=A_{\rm S}(0)=A \, ,
\end{equation}
whereas the states evolve in time according to the Schr\"odinger equation,
\begin{equation}
      \rmi \hbar \frac{\rmd}{\rmd t}\varphi _{\rm S}(t)=
         H\varphi _{\rm S}(t) \, .
\end{equation}
Integration of this equation leads to
\begin{equation}
     \varphi _{\rm S}(t)=\rme ^{-\rmi H t/\hbar}\varphi _{\rm S}(0) \equiv
     \rme ^{-\rmi H t/\hbar}\varphi   \, .
\end{equation}
In the Schr\"odinger picture, the expectation value of the measurement of the 
observable $A$ in the state $\varphi$ is given by
\begin{equation}
     \langle A \rangle  _{\rm S}(t) =  
   \langle \varphi _{\rm S}(t)|A_{\rm S}|\varphi _{\rm S}(t) \rangle =
   \langle \rme ^{-\rmi H t/\hbar}\varphi |A|
       \rme ^{-\rmi H t/\hbar}\varphi \rangle   \, .
      \label{exvaS}
\end{equation}

In the Heisenberg picture, which shall be denoted by the subscript 
${\rm H}$, the states are kept fixed in time, 
\begin{equation}
      \varphi _{\rm H}(t)=\varphi _{\rm H}(0)= \varphi \, ,
\end{equation}
whereas the observables evolve in time according to ``Heisenberg's equation of
motion:''
\begin{equation}
      \rmi \hbar \frac{\rmd}{\rmd t}A_{\rm H}(t)=
         [A_{\rm H}(t), H] \, .
     \label{Heqmot}
\end{equation}
In integrated form, Eq.~(\ref{Heqmot}) reads as
\begin{equation}
       A_{\rm H}(t)=\rme ^{\rmi Ht/\hbar}A_{\rm H}(0)\rme ^{-\rmi Ht/\hbar}
       = \rme ^{\rmi Ht/\hbar}A\rme ^{-\rmi Ht/\hbar}   \, .
     \label{teob}
\end{equation}
In the Heisenberg picture, the expectation value of the measurement of $A$ in 
the state $\varphi$ is given by
\begin{equation}
      \langle A \rangle _{\rm H}(t) = 
    \langle \varphi _{\rm H}|A_{\rm H}(t)|\varphi _{\rm H} \rangle =
   \langle  \varphi | 
    \rme ^{\rmi H t/\hbar}A\rme ^{-\rmi H t/\hbar}|\varphi \rangle  \, .
      \label{exvaH}
\end{equation}
The equivalence of the Schr\"odinger and Heisenberg pictures is guaranteed
by the equality of the expectation values~(\ref{exvaS}) and~(\ref{exvaH}):
\begin{equation}
     \langle A \rangle _{\rm S}(t) =  \langle A \rangle _{\rm H}(t) \, ,
\end{equation}
which follows from the unitarity of the group evolution operator 
$\rme ^{-\rmi H t/\hbar}$.

When the operators of the algebra ${\cal A}$ are all bounded, they are defined
on the whole of the Hilbert space $\cal H$, and domain troubles do not 
arise. But if at least one operator of the algebra, say $A_1$, is unbounded, 
then $A_1$ cannot be defined on the whole of the Hilbert space, but at the 
most on a dense subspace ${\cal D}(A_1)$ of the Hilbert space on which $A_1$ 
is self-adjoint. In such event, one has to specify on
what states the algebraic operations involving unbounded operators are 
valid, since algebraic operations (e.g., sums, products and
commutation relations) of unbounded operators are not defined on the whole 
of ${\cal H}$: If $A$ and $B$ are two unbounded operators defined on two 
dense subdomains ${\cal D}(A)$ and ${\cal D}(B)$ of $\cal H$, then the sum of 
$A$ and $B$, $A+B$, is defined only for 
$f \in {\cal D}(A) \cap {\cal D}(B)$; the product of $A$ by $B$,
$BA$, is defined only for those $f \in {\cal D}(A)$ such that
$Af \in {\cal D}(B)$; the commutation relation of $A$ and $B$, 
$[A,B]= AB-BA$, is defined only for those $f$ such that 
$f \in {\cal D}(A) \cap {\cal D}(B)$, $Af \in {\cal D}(B)$ and 
$Bf \in {\cal D}(A)$.

Likewise algebraic operations, the time evolution~(\ref{teob}) of an unbounded
operator $A$ cannot be defined on the whole of $\cal H$. Clearly,
the time evolution of $A$,
\begin{equation}
       A(t) \equiv  \rme ^{\rmi Ht/\hbar}A\rme ^{-\rmi Ht/\hbar}   \, ,
     \label{teobA}
\end{equation}
is defined only for those $f\in {\cal H}$ such that 
$\rme ^{-\rmi H t/\hbar}f \in {\cal D}(A)$. Thus, the Heisenberg picture of 
an unbounded operator $A$ is not defined on the whole of $\cal H$.

We have therefore to face the fact that algebraic operations and the
Heisenberg picture of unbounded operators entail domain troubles. As we are 
going to see in the next section, such domain troubles can be surmounted by
extending the Hilbert space to the rigged Hilbert space.

\section{Sketch of a solution}

The way the rigged Hilbert space surmounts the domain troubles of algebraic
operations is well known (see~\cite{JPA04} for a recent, simple 
example). Basically,
when resonances are not involved, one has to construct the maximal
invariant subspace of the algebra of operators,
\begin{equation}
      \Phi = \bigcap _{A\in {\cal A}} {\cal D}(A) \, .
\end{equation}
The space $\Phi$ is obviously contained in the domains of the observables of
the algebra,
\begin{equation}
       \Phi \subset {\cal D}(A) \, , \quad  A \in {\cal A} \, ,
\end{equation}
and is the largest subspace of the Hilbert space that remains invariant under 
the action of all the operators of the algebra:
\begin{equation}
     A \Phi \subset \Phi \, , \qquad A \in {\cal A}  \, .
\end{equation}
It is precisely this invariance what makes all algebraic operations (e.g., 
sums, multiplications and commutation relations) well defined on $\Phi$. In
addition, the bras $\langle a|$ and the kets $|a\rangle$ associated with the 
continuous spectra of the operators belong to the dual, 
$\Phi ^{\prime}$, and to the antidual, $\Phi ^{\times}$, spaces, respectively:
\begin{equation}
       \begin{array}{ll}       \langle a| \in  \Phi ^{\prime} \, , \\ [2ex]
       |a\rangle \in  \Phi ^{\times} \, .   
        \end{array} 
\end{equation}

Now, it is clear that in order to avoid the domain troubles of the time 
evolution~(\ref{teobA}) of an unbounded observable $A$,
we simply need to let $A(t)$ act on a subspace whose time evolution is
included in ${\cal D}(A)$. Since we want this to happen for all the operators 
of the algebra, it is natural to demand that the space $\Phi$ be invariant 
under the action of the time evolution group:
\begin{equation}
      \rme ^{-\rmi Ht/\hbar} \Phi \subset \Phi \, .
    \label{invofphte}
\end{equation}
When the invariance~(\ref{invofphte}) holds, the time evolution of all the
operators of the algebra is well defined on $\Phi$, and $\Phi$ is invariant
under $A(t)$:
\begin{equation}
      A(t) \Phi \subset \Phi  \, , \qquad A\in {\cal A} \, .   
\end{equation}
This invariance makes, in particular, all algebraic operations involving
the time evolution of the observables well defined.

The problem is, it is not known whether the invariance~(\ref{invofphte}) 
holds for any Hamiltonian and for any algebra we could 
think of. But, as we are going to see in the next section, for some cases of 
interest a theorem by Hunziker provides a positive answer.

To finish this section we note that, as a byproduct of the 
invariance~(\ref{invofphte}), one can define the time evolution of the bras 
and kets:
\begin{equation}
     \begin{array}{ll} 
      \langle a| (t) = \langle a|\rme ^{\rmi Ht/\hbar} \, , \\ [2ex] 
   |a\rangle (t) =  \rme ^{-\rmi Ht/\hbar} |a\rangle   \, .
     \end{array} 
   \label{tebrasandkets}
\end{equation}
In the rigged Hilbert space language, the precise definition of this time 
evolution is as follows:
\begin{equation}
     \begin{array}{ll} 
     \langle a|\rme ^{\rmi Ht/\hbar}|\varphi \rangle \equiv
     \langle a|\rme ^{-\rmi Ht/\hbar}\varphi \rangle \, , \quad 
      \varphi \in \Phi \, ,  \\ [2ex] 
   \langle \varphi| \rme ^{-\rmi Ht/\hbar} |a\rangle \equiv
    \langle \rme ^{\rmi Ht/\hbar} \varphi| a\rangle  \, , \quad 
      \varphi \in \Phi   \, .
     \end{array} 
   \label{tebrasandketsrhs}
\end{equation}
Because $\langle a|$ and $|a\rangle$ are defined only when they act on 
$\Phi$, definitions~(\ref{tebrasandketsrhs}) make sense only when
$\rme ^{-\rmi Ht/\hbar}\varphi$ and $\rme ^{\rmi Ht/\hbar}\varphi$ belong
to $\Phi$; that is, definitions~(\ref{tebrasandketsrhs}) make sense only when
$\Phi$ is invariant under the time evolution group. Thus, the invariance
of $\Phi$ under the time evolution group, which guarantees that $A(t)$ is well 
defined, also guarantees that the time evolution of the bras and kets is well
defined.

\section{Example}

In practical applications, the most important unbounded observables we 
encounter are the position, the momentum and the energy operators. If, for
simplicity, we restrict ourselves to one dimension, the position
operator $Q$ is defined as multiplication by the position coordinate:
\begin{equation}
      Qf(x)=xf(x) \, ;
\end{equation}
the momentum operator $P$ is defined as differentiation with respect to
the position coordinate:
\begin{equation}
      Pf(x)=- \rmi \hbar \frac{\rmd f(x)}{\rmd x} \, ;
\end{equation}
and the energy operator, or Hamiltonian, $H$ is the sum of the kinetic energy
operator and the potential $V(x)$:
\begin{equation}
      Hf(x)= -\frac{\hbar ^2}{2m}\frac{\rmd ^2f(x)}{\rmd ^2x}+ V(x)f(x) \, .
\end{equation}

As we explained in the previous section, we have to construct the maximal
invariant subspace $\Phi$ of the algebra $\{ H, Q, P \}$, and then see
whether $\Phi$ is invariant under $\rme ^{-\rmi Ht/\hbar}$. The form of
$\Phi$ and the invariance of $\Phi$ under $\rme ^{-\rmi Ht/\hbar}$
depend on the form of $V(x)$. 

In this note, we shall only consider potentials $V(x)$ that are
bounded $C^{\infty}$-functions with bounded derivatives. For such potentials,
the maximal invariant subspace of the algebra $\{ H, Q, P \}$
is the Schwartz space:
\begin{equation}
      \Phi = {\cal S}({R})  \, .
\end{equation}
This space is indeed invariant under $\rme ^{-\rmi Ht/\hbar}$, as the following
theorem states for any dimension $n$~\cite{HUNZIKER}:

\vskip0.3cm

{\bf Theorem} (Hunziker) \ {\it If $V(x)$ is a bounded 
$C^{\infty}$-function on 
${R}^n$ with bounded derivatives, then ${\cal S}({R}^n)$ is invariant under 
the unitary group $\rme ^{-\rmi Ht/\hbar}$ and the mapping 
$(\varphi , t)\to \rme ^{-\rmi Ht/\hbar}\varphi$ of 
${\cal S}({R}^n)\times {R}$ onto ${\cal S}({R}^n)$ 
is continuous (in the sense of the conventional topology of 
${\cal S}({R}^n)$).}

\vskip0.3cm

Thus, when $V(x)$ is a bounded $C^{\infty}$-function with bounded derivatives,
the time evolution (and therefore the Heisenberg picture) of the algebra
$\{ H, Q, P \}$ is well defined on $\Phi = {\cal S}({R})$. As a byproduct, 
the time evolution of the bras  $\langle E|$, $\langle x|$, $\langle p|$ and
kets $|E\rangle$, $|x\rangle$, $|p\rangle$ of $H$, $Q$, $P$ is also well
defined, since those bras and kets act as functionals over
$\Phi = {\cal S}({R})$.

When $V(x)$ is not continuous but only piece-wise continuous (e.g., $V(x)$ is 
a rectangular barrier potential), the discontinuities of the potential have to 
be taken into consideration. For example, if $V(x)$ is continuous everywhere 
except at, say, $x=a,b$, then the maximal invariant subspace of 
$\{ H, Q, P \}$ is the space ${\cal S}({R}-\{ a, b \})$ constructed 
in~\cite{JPA04}. In this case, one has to prove that 
${\cal S}({R}-\{ a, b \})$ is invariant under 
$\rme ^{-\rmi Ht/\hbar}$. Although the invariance of 
${\cal S}({R}-\{ a, b \})$ under $\rme ^{-\rmi Ht/\hbar}$ is still to be 
proven, an examination of the proof of Hunziker's theorem~\cite{HUNZIKER} 
suggests that such invariance should hold.

For other ``reasonable'' potentials the invariance of the corresponding $\Phi$ 
under time evolution is to be expected, too. By ``reasonable'' we mean that
the potential can be considered as a small perturbation to the kinetic 
energy, in the sense of Kato~\cite{KATO}.

\section{Conclusion}

We have discussed the problems in defining the Heisenberg picture of
an algebra of unbounded observables. We have seen that the Heisenberg
picture is well defined when the maximal invariant subspace of the algebra
remains invariant under the time evolution group. Such invariance should
hold in general, although we have shown it only for potentials that
are smooth. More precisely, when $V(x)$ is a bounded $C^{\infty}$-function 
with bounded derivatives, Hunziker's theorem has been used to show that
the Heisenberg picture of the algebra $\{ H, Q, P \}$ is well defined on
the Schwartz space.

\section*{Acknowledgments}

The author is indebted to E.~Galapon for enlightening discussions. Financial
support from the Basque Government through reintegration 
fellowship No.~BCI03.96 is gratefully acknowledged.

{\it Note:} The version of Hunziker's theorem I provided in this paper is
not the most general one. In fact, after this paper was already in press, I 
realized that a theorem by Roberts~\cite{ROBERTS} combined with
the general version of Hunziker's theorem~\cite{HUNZIKER} guarantees that, 
when the potential is infinitely differentiable on some open set of ${R}^n$ 
whose complement has zero Lebesgue measure, and when the potential satisfies
the Kato condition~\cite{KATO}, then the maximal invariant
subspace of the algebra $\{ H, Q, P \}$ is dense and invariant under
$\rme ^{-\rmi Ht/\hbar}$. In particular, the space 
${\cal S}({R}-\{ a, b \})$ of~\cite{JPA04} is indeed invariant under the time 
evolution of the 1D rectangular barrier Hamiltonian.

\end{document}